*Review*

# Rip Cosmology via Inhomogeneous Fluid

**Valerii V. Obukhov, Alexander V. Timoshkin * and Evgenii V. Savushkin**

Department of Physics and Mathematics, Tomsk State Pedagogical University, Tomsk 634061, Russia;
E-Mails: obukhov@tspu.edu.ru (V.V.O.); evgeny-tomsk@rambler.ru (E.V.S.)

* Author to whom correspondence should be addressed; E-Mail: timoshkinav@tspu.edu.ru;
  Tel.: +3-822-996-728; Fax: +3-822-446-826.

**Abstract:** The conditions for the appearance of the Little Rip, Pseudo Rip and Quasi Rip universes in the terms of the parameters in the equation of state of some dark fluid are investigated. Several examples of the Rip cosmologies are investigated.

## 1. Introduction

The appearance of new cosmological models is connected with the discovery of the accelerated expansion of the universe. Cosmic acceleration can be introduced via dark energy [1,2] or via modification of gravity [3,4]. Dark energy should have strong negative pressure and can be characterized by an equation of state (EoS) parameter $w$. The thermodynamic parameter $w = p/\rho$, where $\rho$ is the dark energy and $p$ is the dark pressure, is known to be negative. The theory predicts many interesting ways in which the universe could have evolved, including the Big Rip (BR) [5,6], the Little Rip (LR) [7–14], the Pseudo Rip (PR) [15] and Quasi Rip (QR) [16] cosmological models. The BR singularity phenomenon means that the physical quantities become infinite at the finite Rip time $t$. In the LR scenario, an infinite time is required to reach the singularity. In the PR cosmology, the Hubble parameter tends to the "cosmological constant" in the remote future. In the Rip phenomena, like LR or PR, the parameter $w$ asymptotically tends to $-1$. These models are based on the assumption that the dark energy density $\rho$ is a monotonically increasing function. In the cosmological model, the QR the dark energy density $\rho$ monotonically increases when EoS parameter $w < -1$ in the first stage, and monotonically decreases ($w > -1$) in the second stage.

In this review article we study the influence of the time-dependent thermodynamic parameter *w* and the cosmological constant Λ from the EoS on the occurrence the Rip phenomena of some cosmological models. Section 2 is devoted the non-viscous models of the cosmic fluid. In Section 3 we consider the description of viscous LR cosmology for dark fluid in the late universe.

## 2. Dark Fluid Inhomogeneous Equation of State in the Some Cosmological Models

We suppose that our universe is filled with an ideal fluid (dark energy) obeying an inhomogeneous EoS [17]:

$$p = w(t)\rho + \Lambda(t) \tag{1}$$

where *p* is the pressure and *w*(*t*), Λ(*t*) are time-dependent parameters.

The Friedmann equation for a spatially flat universe is:

$$\rho = \frac{3}{k^2}H^2 \tag{2}$$

where ρ is the energy density, $H = \dot{a}/a$ is the Hubble parameter, *a*(*t*) is the scale factor, $k^2 = 8\pi G$ with Newton's gravitational constant *G*.

Let us write down the energy conservation law:

$$\dot{\rho} + 3H(p+\rho) = 0 \tag{3}$$

We now consider examples of the dark energy models corresponding to the LR, QR and PR universes. For simplicity it will be assumed that the universe consists of the dark energy only.

*2.1. The Little Rip Case*

Let us a Hubble parameter has the following form [9]:

$$H(t) = H_0 e^{\lambda t} \tag{4}$$

where $H_0 > 0$, $\lambda > 0$, and $H_0$ is the present-time Hubble parameter.

We assume that the parameter *w* does not depend on the time $w(t) = w_0$.

Taking into account Equations (1)–(4) and solve the Equation (3) with respect to Λ(*t*), we obtain [18]:

$$\Lambda(t) = -\frac{H}{k^2}\left[2\lambda + 3H(1+w_0)\right] \tag{5}$$

Thus, if we assume an ideal fluid obeying the EoS Equations (1) and (5), then we obtain the LR scenario.

Let us consider another LR model [9]:

$$H(t) = H_0 e^{Ce^{\lambda t}} \tag{6}$$

Here, $H_0$, *C* and λ are positive constants.

Now writing the parameter *w*(*t*) in the form:

$$w(t) = -1 - \frac{\delta}{H^2},\ \delta > 0 \tag{7}$$

and express Λ(*t*) from Equation (3) we obtain the solution, realizing the LR (6) [18]:

$$\Lambda(t) = -\frac{2\lambda}{k^2}\ln\frac{H}{H_0} + \frac{3\delta}{k^2} \qquad (8)$$

Let us choose the cosmological model with more complicated behavior of *H* [9]:

$$H = H_0 \exp\left(C_0 \exp C_1 \left(\exp C_2 \left(\exp\ldots\left(C_n \exp(\lambda t)\right)\right)\right)\right) \qquad (9)$$

where, $C_0$, $C_1$, …, $C_n$, are the positive constants, and use Equations (6) and (7) for the solution Equation (3) with respect to Λ(*t*). By generalizing Equation (8), we obtain [18]:

$$\Lambda(t) = -\frac{2\lambda}{k^2} H \ln\frac{H}{H_0} Ln\frac{H}{H_0} + \frac{3\delta}{k^2} \qquad (10)$$

Here we have defined $L_n$ as:

$$Ln\frac{H}{H_0} \equiv \left[\ln\left(\frac{1}{C_0}\ln\frac{H}{H_0}\right)\ldots\ln\left(\frac{1}{C_{k-1}}\ldots\ln\left(\frac{1}{C_0}\ln\frac{H}{H_0}\right)\right)\right] \qquad (11)$$

We consider also the example of the brane LR cosmology [12]. The Hubble parameter is equal [19]:

$$H = \sqrt{\frac{\lambda}{6}} sh\left(\sqrt{\frac{3}{2\lambda}}\alpha^2 t\right) \qquad (12)$$

where λ is a positive tension (λ > 0).

Let us choose the parameter *w*(*t*) as:

$$w(t) = -1 - \frac{2}{\lambda sh^2\left(\sqrt{\frac{3}{2\lambda}}\alpha^2 t\right)} \qquad (13)$$

then we find in analogy the "cosmological constant" [19]:

$$\Lambda(t) = 1 - \alpha^2 ch\left(\sqrt{\frac{3}{2\lambda}}\alpha^2 t\right) \qquad (14)$$

As result, we have obtained a brane dark energy universe from the standpoint of 4d FRW cosmology without introducing the brane conception.

*2.2. The Pseudo Rip Case*

Let us investigate a PR model with the parameter Hubble [9]:

$$H(t) = H_0 - H_1 e^{-\lambda t} \qquad (15)$$

where $H_0$, $H_1$ and λ are the positive constants. We assume that $H_0 > H_1$ when *t* > 0.

If $t \to +\infty$ the Hubble ratio tends to a constant value $H_0$ and the universe asymptotically approaches the de Sitter space. It may correspond to a PR model.

We will consider this cosmological model in analogy with the LR model.

Let us take the parameter *w*(*t*) in the view Equation (7), we obtain [18]:

$$\Lambda(t) = -\frac{2\lambda}{k^2}(H - H_0) - \frac{3}{k^2}\delta \qquad (16)$$

showing that the PR behavior is caused by the parameter *w*.

In the next example we have investigated the appearance of the asymptotic de Sitter regime on the brane from 4d cosmology [12]. The Hubble parameter is equal [19]:

$$H = \frac{\alpha\beta}{2} \cdot \frac{tg\eta_0 + \frac{\alpha}{\beta}t}{\sqrt{1 + \left(tg\eta_0 + \frac{\alpha}{\beta}t\right)^2}} \qquad (17)$$

Here $\beta^2 = \frac{2|\lambda|}{3\alpha^2}$ is the dimensionless parameter, $\eta_0 = \sqrt{\frac{3}{2|\lambda|}}\alpha^2 t_0$, where $t_0$ is the present time and λ is a negative tension (λ < 0). If $t \to \infty$, then the Hubble parameter $H \to \frac{\alpha\beta}{2}$. This situation corresponds to the universe expands in a quasi-de Sitter regime.

Now writing the parameter *w*(*t*) in the view:

$$w(t) = -1 - \frac{\delta}{3H} \qquad (18)$$

with δ > 0, we obtain [19]:

$$\Lambda(t) = -H\left(\frac{2\frac{\alpha}{\beta}}{tg\eta_0 + \frac{\alpha}{\beta}t} - \frac{\delta}{3H}\right) \qquad (19)$$

where $t \neq -\frac{\beta}{\alpha}tg\eta_0$.

*2.3. The Quasi Rip Case*

In this case we have modeled the QR universe induced by the dark fluid EoS. Let us take the energy density as a function of the scale factor *a* [14]:

$$\rho = \rho_0 a^{\alpha - \beta \ln a} \qquad (20)$$

where α and β are a constants, $\rho_0$ is the energy density at a present time $t_0$. Now we write the EoS parameters *w*(*a*) and Λ(*a*) depending on the scale factor *a*.

Choosing the parameter *w*(*a*) in the EoS as:

$$w(a) = -1 - \frac{\delta}{3\rho_0}a^{\beta \ln a - \alpha} \qquad (21)$$

where δ is a constant, we obtain the solution realizing QR Equation (20) caused by the parameter *w*(*a*) [20]:

$$\Lambda(a) = -\rho_0 a^{\alpha - \beta \ln a}(\alpha - 2\beta \ln a) - \frac{\delta}{3} \qquad (22)$$

## 3. Examples of the Viscous Little Rip Cosmology

In this section we will consider the examples of the viscous LR cosmology in an isotropic cosmic fluid in the later stage of the evolution of the universe.

*3.1. Dark Fluid with Bulk Viscosity*

Let us write the expression for the time-dependent energy density for the viscous LR cosmology [21]:

$$\rho(t) = \left[\left(\frac{\xi_0}{A} + \sqrt{\rho_0}\right)\exp\left(\sqrt{6\pi G}At\right) - \frac{\xi_0}{A}\right]^2 \quad (23)$$

with the viscous condition $\xi_0 = 3\zeta H = const$, where $\zeta$ is the bulk viscosity and $A$ is a positive constant. Now we consider the viscous LR cosmology from the point of view of 4d FRW non-viscous cosmology, analogous to Section 2.

If the parameter $w(t)$ has the form:

$$w(t) = -1 - \frac{\delta}{3AH^2} \quad (24)$$

with $\delta > 0$, then the "cosmological constant" is equal [22]:

$$\Lambda(t) = \delta - 2\sqrt{2\pi G}\left(\sqrt{3}AH + \xi_0\right) \quad (25)$$

The validity of the Equations (24) and (25) means an equivalent description the viscous LR (23).

*3.2. The Turbulent Description*

In the later stages of the evolution of the universe near the future singularity it is necessary to take into account a transition into the turbulence motion. Let us consider the cosmic fluid as a two-component fluid and introduce the effective energy density in the view [21]:

$$\rho_{eff} = \rho + \rho_{turb} \quad (26)$$

The first term $\rho$ denotes the ordinary laminar energy density and the second term $\rho_{turb}$ denotes the turbulent part. Now we present analogously the effective pressure:

$$p_{eff} = p + p_{turb} \quad (27)$$

The dependence of $p$ on $\rho$ is given with simple relation:

$$p = w\rho \quad (28)$$

Analogously the turbulent quantities $p_{turb}$ and $\rho_{turb}$ are connected by the similar form:

$$p_{turb} = w_{turb}\rho_{turb} \quad (29)$$

where $w_{turb}$ is a constant.

Let us consider the case $w_{turb} = w < -1$, that is the turbulent matter behaves similar the non-turbulence matter in the phantom region. We will investigate the LR model and take the effective energy density as [21]:

$$\rho_{\textit{eff}} = \frac{\xi_0^2}{9A^2}\left[1+\left(\frac{3A\sqrt{\rho_0}}{\xi_0}-1\right)\exp\left(\frac{1}{2}\sqrt{3}At\right)\right]^2 \quad (30)$$

The viscous LR model for a perfect fluid can be realized via the choice in the EoS the parameter $w(t)$ in the view:

$$w(t) = -1 - \frac{\delta k^2}{3H^2} \quad (31)$$

and corresponding expression for $\Lambda(t)$ [22]:

$$\Lambda(t) = \frac{\xi_0}{3} + \delta - \frac{\sqrt{3}}{k}AH \quad (32)$$

Note, that there is another method of the solving this problem, which is connected with the transition of a one-component cosmic fluid from the viscous era into the turbulent era [21].

## 4. Conclusions

Several dark energy models have been analyzed in the present review article. We showed that these cosmological models can be caused via the corresponding choice of the cosmological constant or the thermodynamic parameter in the dark fluid inhomogeneous EoS within the framework of 4d FRW cosmology.

## Acknowledgements

This work has been supported by project 2.1839.2011 of Ministry of Education and Science (Russia) and LRSS project 224.2012.2 (Russia). We are very grateful to Professor Sergei Odintsov for helpful discussions.

## Conflicts of Interest

The authors declare no conflict of interest.